\documentclass[prb,twocolumn,superscriptaddress]{revtex4}
\usepackage{amsfonts}
\usepackage{amsmath}
\usepackage{amssymb}
\usepackage{graphicx}
\usepackage{hyperref}

\usepackage{braket}

\begin{document}

\title{Role of Entropy and Structural Parameters in the Spin State Transition of LaCoO$_3$}
\author{Bismayan Chakrabarti}
\affiliation{Department of Physics \& Astronomy, Rutgers University, Piscataway, NJ 08854-8019, USA}
\author{Turan Birol}
\affiliation{Department of Physics \& Astronomy, Rutgers University, Piscataway, NJ 08854-8019, USA}
\affiliation{Department of Chemical Engineering and Materials Science, University of Minnesota, Minneapolis, MN 55455-0132, USA}
\author{Kristjan Haule}
\affiliation{Department of Physics \& Astronomy, Rutgers University, Piscataway, NJ 08854-8019, USA}

\date{\today}

\begin{abstract}
The spin state transition in LaCoO$_3$ has eluded description for decades despite concerted theoretical and experimental effort. In this study, we approach this problem using fully charge self-consistent Density Functional Theory + Embedded Dynamical Mean Field Theory (DFT+DMFT). We show from first principles that LaCoO$_3$ cannot be described by a single, pure spin state at any temperature. Instead, we observe a gradual change in the population of higher spin multiplets with increasing temperature, with the high spin multiplets being excited at the onset of the spin state transition followed by the intermediate spin multiplets being excited at the metal insulator transition temperature. We explicitly elucidate the critical role of lattice expansion and oxygen octahedral rotations in the spin state transition. We also reproduce, from first principles, that the spin state transition and the metal-insulator transition in LaCoO$_3$ occur at different temperature scales. In addition, our results shed light on the importance of electronic entropy in driving the spin state transition, which has so far been ignored in all first principles studies of this material. 
\end{abstract}
\maketitle

\section{Introduction}

The spin state transition in LaCoO$_3$ has been the subject of  intensive investigation for decades.\cite{Heikes, Naiman, Goodenough} This compound is established to be a narrow bandgap insulator at low temperature with Pauli-like magnetic susceptibility. However between 90-150 K, it transitions to a local moment phase with a Curie-Weiss like susceptibility which reaches its peak around 150K. It also undergoes a gradual closing of the insulating gap and is known to be metallic above 600K \cite{BhideMoss, English, Saitoh}.

There is considerable debate regarding the mechanism of this transition, mainly due to the uncertainty regarding the multiplet of the $Co^{3+}$ ion which  characterizes the excited state of the compound.  The cobalt ion in LaCoO$_3$ is commonly assumed to be in the $d^6$ state with a formal valence of $3+$.  Due to the fact that the scale of the crystal field splitting is comparable to the Hunds coupling energy scale, one would expect that as temperature is increased, there would be an entropy-driven transition from the low spin (LS) $S=0$ state with a fully filled $t_{2g}$ shell($t^6e^0$) to an $S=2$ high spin (HS) state ($t^4e^2$)\cite{Goodenough}. Indeed there is considerable experimental evidence to support such a scenerio. Electron spin resonance\cite{Zopka}, neutron scattering \cite{Podlesnyak}, X-ray absorbtion spectroscopy and magnetic circular dichroism experiments\cite{Haverfort} all point towards a transition to an HS state. In addition, no inequivalent Co-O bond is found in EXAFS experiements, which also supports the formation of an HS state due to the HS state not being strongly Jahn-Teller active \cite{Sundaram}. However, it has been noted that in order to explain the  XAS experimental data, one would have to assume that the crystal field grows with temperature, which is counter-intuitive.\cite{Eder, Haverfort} This led to some authors suggesting that there is an LS-HS alternating structure caused by breathing distortions\cite{Bari} \cite{Goodenough} and interatomic repulsion between the HS atoms.\cite{Asaka, Eder} 
 
A competing explanation, whereby the excited state is the $S=1$ intermediate spin (IS) state ($t^5e^1$), has also become popular\cite{Heikes, Radaelli}, mainly because of LDA+U results which show that the    
IS state is lower in energy compared to the HS state.\cite{Korotin, Pandey, Anisimov} The stability of the IS state has been justified by the large hybridization of the Co 3d electrons with neighboring O 2p electrons. This causes charge transfer between the ions resulting in the Co ion having a $d^7$ structure according to the Zaanen-Sawatzky-Allen scheme,\cite{Zaanen} which in turn would cause stabilization of the IS state. The intermediate spin state hypothesis also seems to explain experimental findings such as Raman Spectroscopy, X-Ray photoemission, XAS, EELS, as well as susceptibility and thermal expansion measurements. \cite{Saitoh, Abbate, Masuda, Klie, Zobel, Gne, Maris, Vogt}. 

To summarize, there has been significant debate regarding the true nature of the spin state transition in LaCoO$_3$. Interest in this compound has also been enhanced in light of recent discoveries of ferromagnetism induced by Sr (hole) doping \cite{Kriener, Masayuki, Kunes_doped,nemeth}, and by experiments reporting strain induced magnetism in epitaxially grown thin films.\cite{Fuchs1, Fuchs2, Rondinelli, Freeland, Fuchs3, Herklotz,Hsu} Additionally, there have been reports of the emergence of a striped phase in thin films (with alternating LS and HS/IS regions)\cite{Striped}, as well as the presence  of low temperature ferromagnetism  in experiments performed on LaCoO$_3$ nanoparticles\cite{Belanger1}. Furthermore,  there have also been experiments conducted on single crystals of LaCoO$_3$ in the presence of a strong magnetic field\cite{highfield1,highfield2,highfield3,highfield4} which have reported the presence of  multiple metamagnetic transitions, which have been attempted to be explained by spin-state superlattices as well as excitonic condensation.\cite{kunes_highfield} Hence, we can see that there is great interest in understanding the complex physical nature of this material. 

There have been multiple previous DMFT studies that have approached this material and the transitions therein. For example, in Ref. \onlinecite{Kunes_doped}, Augustinski et al. studied the effect of carrier doping on the spin state of LaCoO$_3$. Zhang et al, in Ref. \onlinecite{Zhang}, considered hydrostatic pressure as well as substitution of La with other lanthanides, and provided an explicit picture of the effect of the crystal field splitting on the Co ion, while Krapek et al. (Ref. \onlinecite{Kunes_main}) showed that the charge fluctuations, in addition to the spin state fluctuations, are important in understanding the physics of this compound. In this paper, we use Density Functional Theory + embedded Dynamical Mean Field Theory (DFT+DMFT)\cite{dmft_website} to analyze the spin state transition in bulk LaCoO$_3$.  Unlike earlier studies, our implementation is fully charge self-consistent and extremizes the DFT+DMFT functional in real space (for details the reader is referred to the appendix and the references contained therein), thereby avoiding the downfolding approximation and uses the numerically exact CTQMC impurity solver\cite{Dmft3,haule}.  In addition, to our knowledge none of the earlier studies provide a comprehensive analysis of all of the factors governing the transition such as octahedral rotations and electronic entropy. Our main findings can be summarized as- i) We show that LaCoO$_3$ has large charge fluctuations and it is not possible to explain the spin state with a single multiplet at any temperature. However at the onset of the spin state transition, the HS multiplets are excited, with the IS multiplets being excited later around the onset of the metal-insulator transition.   ii) We illustrate that the crystal field splitting is very sensitive to the crystal structure, and taking into account not only the 
thermal expansion but also the oxygen octahedral rotations are very important for understanding the behavior of the material.  We repeat 
our calculations for four different crystal structures that correspond to two different temperatures and two different 
oxygen octahedral rotation angles. This way, for the first time, we isolate the effect of the oxygen octahedral 
rotations  and thermal expansion on the electronic and spin structure of LaCoO$_3$. This should be an important 
factor in determining the effect of Ln substitution, as discussed in Ref. \onlinecite{Zhang} and Ref. \onlinecite{ topsakal2016}, 
but this connection 
has not been studied explicitly in LaCoO$_3$ yet. iii) We demonstrate conclusively that it is possible to stabilize (without orbital order) an insulating phase  at intermediate temperatures where local moments are present, thereby showing that the metal-insulator transition is distinct from the spin state transition in this compound.We provide a detailed picture of the evolution of the spectral function and the spin state probabilities with temperature  and conclusively show from first principles that the spin 
state and metal insulator transitions occur at very different temperatures.   iv)We also show that electronic entropy difference between the high and low temperature states is necessary for the stabilization of the excited spin states, which is a fact that has, though expected to be important, been overlooked in various first principle studies so far primarily because accurate first principles methods for calculation of the electronic entropy have not been available.

\section{Crystal Structure}

\begin{figure}[htbp]
 \begin{center}
 \includegraphics[height=2in]{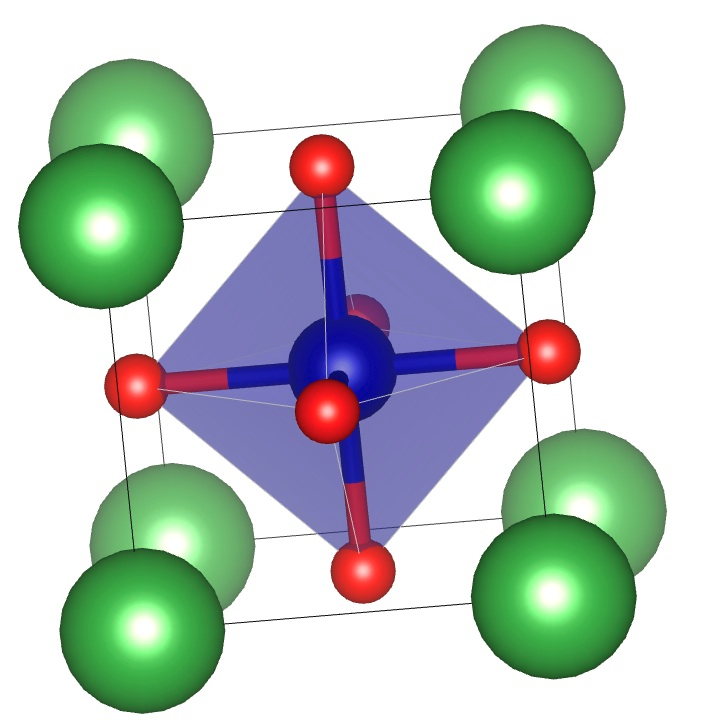}
 \caption{Unit cell for a perovskite with no octahedral rotations ($a^0a^0a^0$ structure). In LaCoO$_3$, the green atoms would be 
La, the red atoms O and the blue atom Co. The figure also shows the octahedra formed by the oxygen atoms around the Co atom.}\label{Perov_struct}
\end{center}
\end{figure}

\begin{figure}[t]
 \begin{center}
 \includegraphics[height=2in]{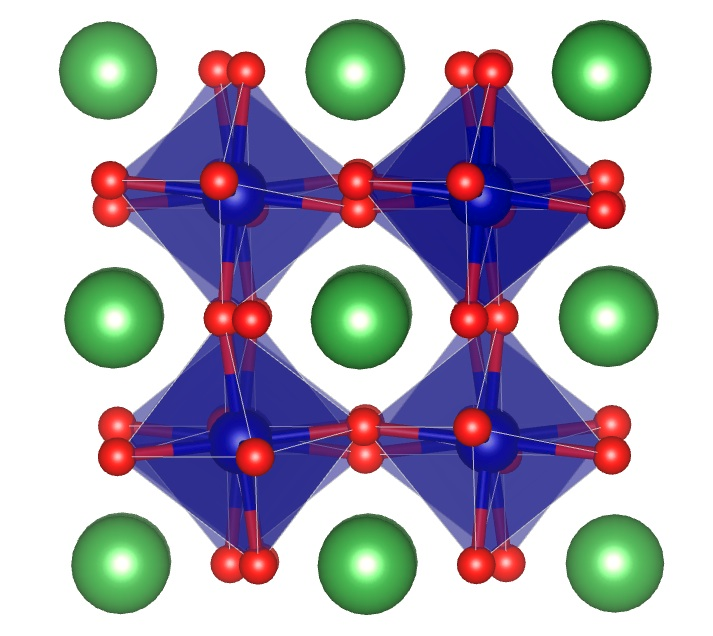}
 \caption{A depiction of the pattern of octahedral rotations that is present in LaCoO$_3$. Each of the oxygen octadra rotates in opposite direction to all nearest neighbour octahedra by the same amount relative to all three cartesian axes ($a^-a^-a^-$ structure).}\label{structure}
 \end{center}
 \end{figure}

LaCoO$_3$ is a perovskite, which has the rare earth element La on the A-site and Co on the B-site at the center of an oxygen octahedron (see Fig \ref{Perov_struct}). Like most perovskites, LaCoO$_3$ has oxygen octahedral rotations (Fig. \ref{structure}) which result in the oxygen octahedra rotating out-of-phase around the [111] axes of the undistorted cubic highsymmetry structure. This rotation pattern is denoted by $a^-a^-a^-$ in  the Glazer notation, and corresponds to the space group R$\bar 3$c (\#167).

As noted by Thornton et. al.\cite{Thornton}, LaCoO$_3$ shows large thermal expansion as well as  variation in the  octahedral rotation angle with increase in temperature. In our study, we use four different crystal structures to isolate and study the effect of different lattice parameters on the spin state transition. We use two different experimental structures observed at 1143K and 4K, which we denote by HTa$^-$ and LTa$^-$. Comparing the electronic structure for these two crystal structures provides a means to study the temperature evolution of the electronic structure. In addition to these two, we also built two crystal structures with the same strain state (unit cell vectors) as them, but with no octahedral rotations. These structures, denoted by HTa$^0$ and LTa$^0$, enable us to isolate the effect of oxygen octahedral rotations on the spin state of LaCoO$_3$. 

\section{Density of States}

\begin{figure*}[t]
\begin{center}
\includegraphics[width=\textwidth,height=3.3in]{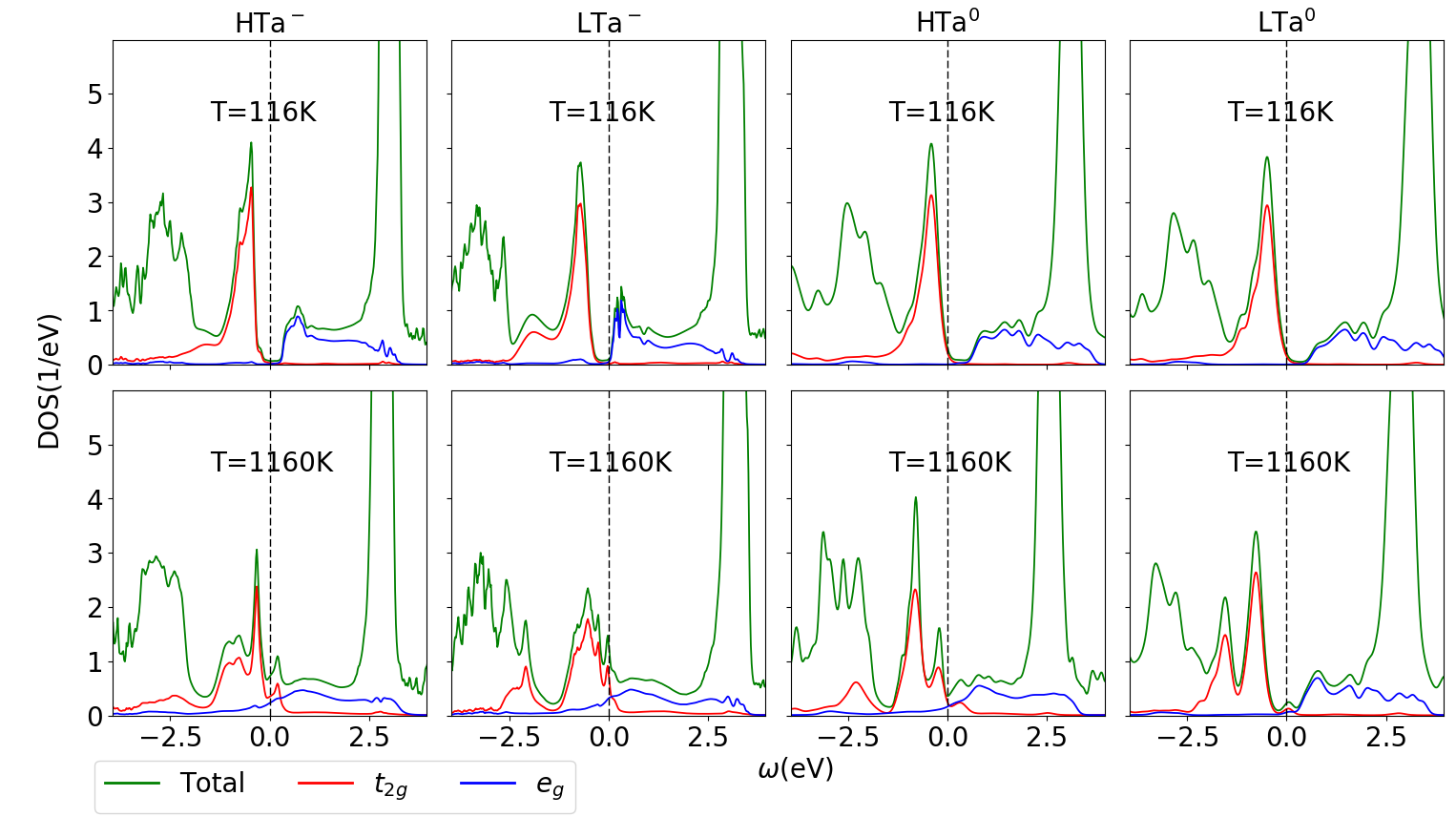}
\caption{Density of states ($t_{2g}$,$e_g$ and total including all atoms) for all four structures at 116K and 1160K.} \label{dos_fig}
\end{center}
\end{figure*}

\begin{figure}[t]
\begin{center}
\includegraphics[width=\columnwidth]{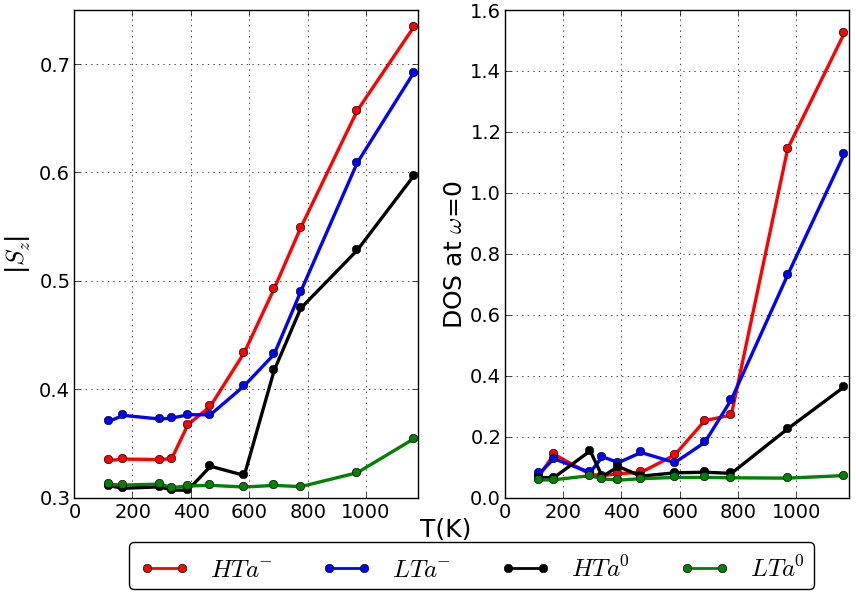}
\caption{(a) Evolution of $|S_z|$ with temperature for all four structures. (b) Evolution of Density of states at fermi level with temperature for all four structures.}\label{fig4}
\end{center}
\end{figure}

In Fig \ref{dos_fig} we show the density of states for all 4 structures, calculated at both low temperature and high temperature (116K and 1160K) using DFT+DMFT. In our DMFT calculations, we use $U=6.0$ eV and $J=0.7$ eV. Further details of our calculations and methodology are presented in the appendix.  Unlike DFT, which always predicts a metallic state, our calculations correctly reproduce an insulating ground state at low temperature for all the structures. The $t_{2g}$ orbitals are below the fermi level whereas the $e_g$ orbitals are above the fermi level. For the two experimental structures (LTa$^-$ and HTa$^-$), this low temperature charge gap closes continuously with increasing temperature and there is a large overlap in energy between the $t_{2g}$ and $e_g$ orbitals at high temperatures. This overlap, however, is much smaller if the structures without rotations are used. (See Fig. \ref{fig4}b).\footnote{While calculating the density of states (DOS) at the Fermi level for the different structures at different temperatures, we ensured that their Fermi energies were adjusted such that the energy levels for the oxygen densities of states lay at the same energy values. This was required because there was an ambiguity in the value of the chemical potential at temperatures where the structure gave rise to an insulating band-gap and we believe an accurate comparison can only be made if some features of the DOS are held fixed. This procedure required a shift in the chemical potential of some of the simulations of the order of 0.1 eV. The results we plot in Fig. \ref{fig4}b are obtained after these shifts are put in. Fig 3 on the other hand plots the densities of states before any such post-processing has been done. This leads to small differences between the two figures.  Instead of fixing the Oxygen levels, we also tried fixing the Lanthanum f levels and this gave rise to very similar results. We firmly believe that  our results displayed in Fig. 4b are robust and it is merely the relevant magnitudes of the y-axis values at high temperatures that fluctuate by a small amount (depending on which features are held fixed) and not the actual temperature at which the charge-gap closure takes place. We also do not plot the DOS at the Fermi level but the average of the DOS at five points around $E=E_F$ as this takes care of some of the numerical noise that creeps into our calculation due to both Monte Carlo noise and the errors in analytic continuation. We also tested our results by averaging over different number of points and no significant changes take place that would affect our claims.}
The HTa$^0$ structure  shows some overlap at high temperatures, while the LTa$^0$ structure almost remains an insulator for the entire range of temperatures studied, with a small overlap developing above 900K. 
This shows clearly that octahedral rotations play a large role in decreasing the strength of the crystal field splitting. 
This can be explained by the fact that the rotation of the oxygen octahedra causes misalignment of the crystal field of the O atoms with that of the La atoms, which normally reinforce each other if the  perovskite has no octahedral rotations. This leads to an overall reduction of the effective crystal field which reduces the charge gap between the $t_{2g}$ and $e_g$ orbitals. \footnote{In addition, this trigonal distortion also leads to a splitting of the $t_{2g}$ orbitals into 2+1 orbitals, thereby again reducing the gap with the $e_g$ orbitals. However this effect is very small and hence is not shown in our plots for the sake of simplicity}. This effect seems to overcome the expected decrease in the bandwith of the $e_g$ orbitals caused by the octahedral rotations. Finally, note that there is a considerable overlap in energy of the O 2p orbitals with the Co 3d orbitals, which is very important in producing charge fluctuations on the Co ion, making it highly mixed-valent. 

\section{The spin state transition}
 
In order to focus on the spin state of the Co ion, we calculate the expectation value of the magnitude of z-component of the spin $\langle|S_z|\rangle$. Note that all our calculations are in the paramagnetic state and hence the value of $\langle S_z \rangle=0$. The results are presented in Fig. \ref{fig4}a as a function of temperature.\footnote{Note that the quantitative value of the transition temperature is overestimated in our calculations. This can be explained by the fact that DMFT does not take into account finite wavelength fluctuations, and as a result, has a tendency to overestimate order like many other mean field methods.} 
The largest value of $|S_z|$ at 1160K is seen for the HTa$^-$ structure, followed by the LTa$^-$ structure. 
This is in line with the stronger crystal field in the LTa$^-$ structure due to the smaller lattice constant.
We also observe that  the spin state transition starts at a higher temperature for the LTa$^-$ structure ($\sim$580K) compared to the the HTa$^-$ ($\sim$380K). This is also consistent with the the low temperature structure having higher stability for the LS state. 
The structures without rotations consistently show lower buildup of higher spin states than the ones with rotations. The HTa$^0$ structure displays a spin state transition, but with an eventual high temperature value of $|S_z|$ that is lower than both the structures with octahedral rotations (LTa$^-$ and HTa$^-$). On the other hand, the LTa$^0$ structure shows almost no transition. This shows that the role that the octahedral rotations play in the reduction of the crystal field is essential for the spin state transition.

Figures \ref{fig4}a and \ref{fig4}b also show that the spin state transition and the charge gap closing occur at different temperatures, which is a trend that has been observed in experiment but has not been captured in earlier DMFT simulations. For example, Fig \ref{fig4}b shows that both the HTa$^-$ and the LTa$^-$ structures show a complete closure of the charge gap at $\sim$ 600K whereas Fig 4a shows that the spin state transition in the two structures occurs at very different temperatures.

\section{Nature of the excited spin state}

Because of the large hybridization between Co and O ions, the Co-d orbitals have large charge fluctuations and all the four structures have an effective d-shell occupation of $n_d \sim 6.6$. As a result, any analysis of the spin states in terms of the LS, IS and HS states of the $d^6$ configuration of the Co ion is necessarily inadequate. In fact, our calculations show that the $d^7$ configuration has a higher occupation probability than $d^6$, and there are also significant probabilities for $d^5$ and $d^8$. 
(Importance of large charge fluctuations in LaCoO$_3$ has been discussed before, for example by Abbate et al.\cite{Abbate} who emphasized the role of covalency between the Co d orbitals and the O anions in order to explain their XAS data.  Our results agree with their observation of highly covalent states and gradual transitions, and show that both the  $3d^6$ and $3d^7$ occupancies are fundamental to understanding the behavior of the material.)

\begin{figure}[h!]
 \begin{center}
 \includegraphics[width=\columnwidth]{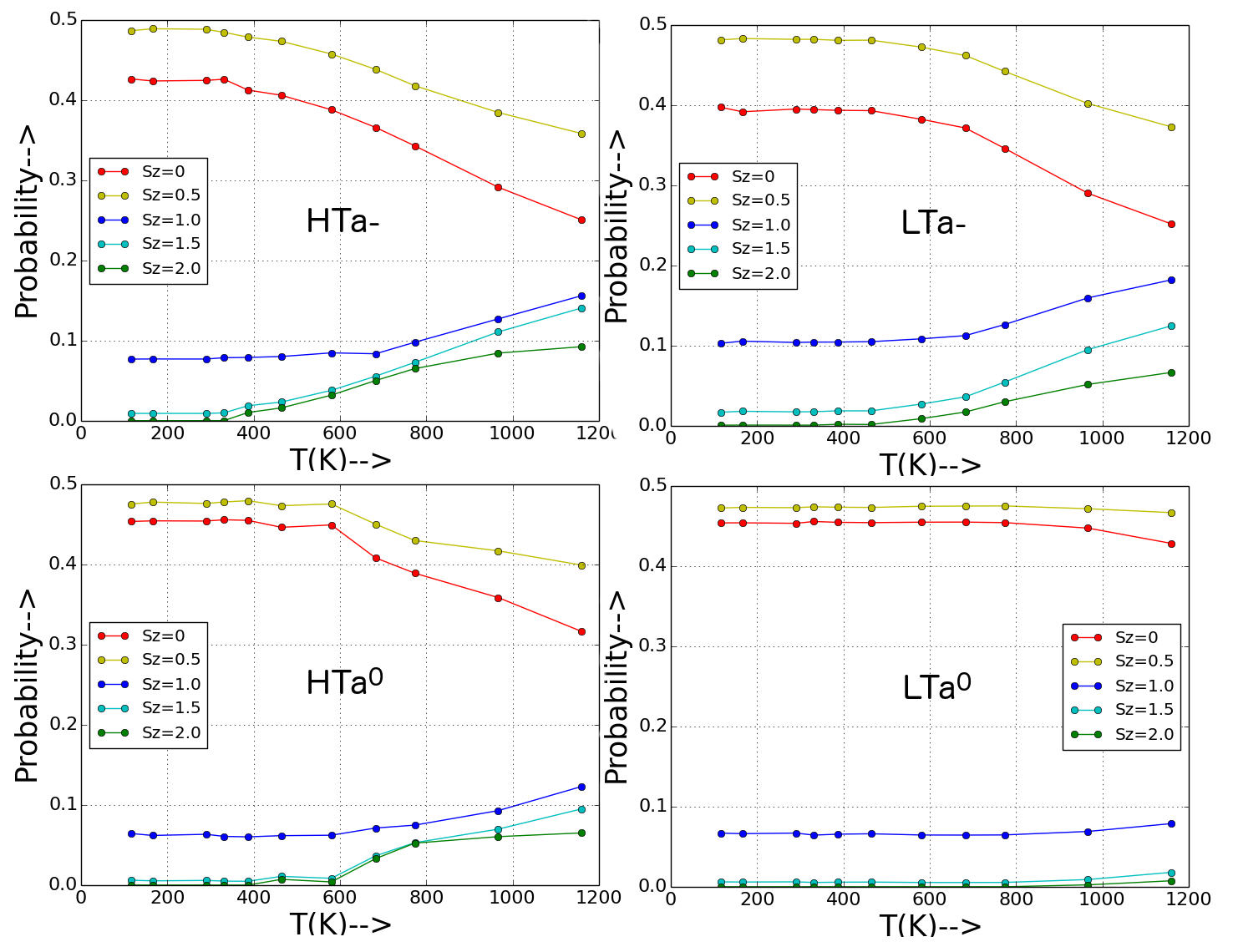}
 \caption{Evolution of occupation probabilities for all the spin states for the four structures with temperature. }\label{spin_prob}
 \end{center}
 \end{figure}

Fig. \ref{spin_prob} shows the evolution of the occupation probabilities for the different values of $|S_z|$ with temperature. Even at high temperatures, $|S_z|=0$ and $|S_z|=0.5$ (the LS states for the even and odd occupancy sectors of the d orbital) remain the states with the highest probability. However, with the increase of temperature, the weight of the higher spin states increases. 
 At the onset of the transition, the initial change in the value of the spin state is predominantly caused by the excitation of the $|S_z|=2$ and the $|S_z|=1.5$ multiplets. For example, from Fig \ref{fig4}a, we can see that for the HTa$^-$ structure the $|S_z|$ value first starts rising at around 300K. Looking at the relevant portion of Fig 5, we see that the only spin states which show an increase in probability are the $|S_z|=2$ and the $|S_z|=1.5$ multiplets.  The $|S_z|=1$ multiplet sees an increase in probability at higher temperatures (above 600K) which is the temperature of the charge gap closing, as evidenced by Fig \ref{fig4}b. A similar trend is seen for the other structures as well. Therefore the initial signature of the transition is best seen in the behavior of the $|S_z|=2.0$ and $|S_z|=1.5$ multiplets, which can be said to be the HS multiplets for the $d^6$ and $d^7$ occupancies respectively, whereas the  $|S_z|=1$ state sees in increase in occupation at the temperature scale of  the metal-insulator transition in each of the structures and not at the spin state transition temperature. (Note that these effects are not seen in the LTa$^0$ structure where no significant transition occurs.)

\begin{figure*}[htbp]
\includegraphics[width=\textwidth,height=3.3in]{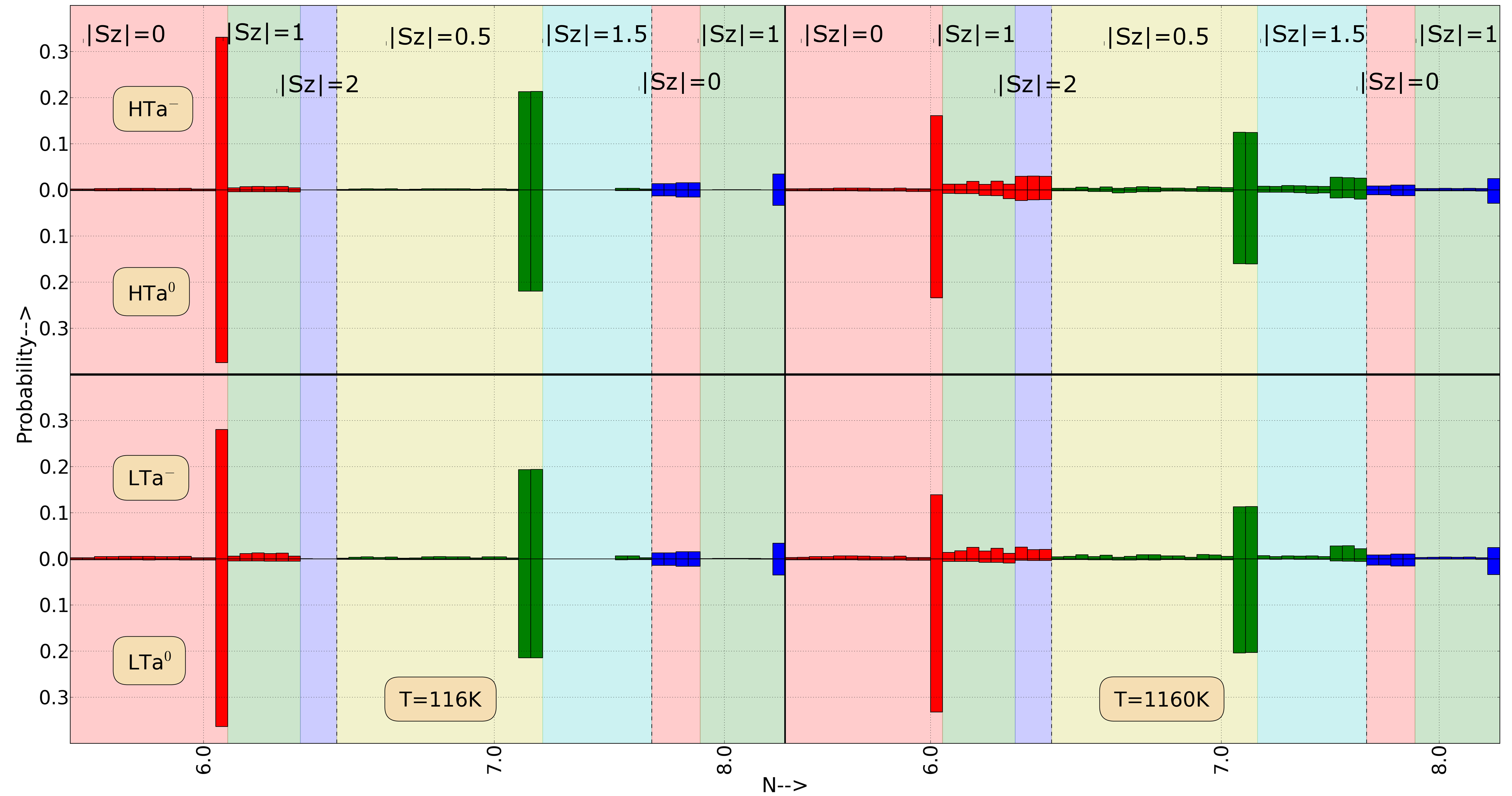}
\caption{Occupancy histogram showing the occupancy probability for the different atomic states for the d orbital of the Co atom for all the four structures at two different temperatures. The different background colors mark the areas reserved for different spin sectors.  The lower x axis ticks as well as the color of the histogram bars denote the different occupancies of the d orbital. Note that odd occupancies only allow half integer values of $|S_z|$ while even ones allow integer values}\label{hist}
\end{figure*}

In Fig. \ref{hist}, we show the occupancy histograms below and above the transition (at 116K and 1160K) (CTQMC gives us access to the state space probability for each of the 1024 states of the d orbital. However, in order to aid visualization, we only show states which have an occupation probability above 0.001 in any of the structures at any temperature). This figure displays clearly how the transition is marked by the excitation of states in the higher spin multiplets. We see that the low temperature state for all of the structures is marked by the presence of a few states with large probability (mainly corresponding to the $|S_z|=0$ and $|S_z|=0.5$ states). As the spin state transition sets in, a large number of higher spin states get excited and the LS spin states lose weight. Note that the high spin states are highly degenerate so there is no one large peak for the high spin states, but instead there is a multitude of lower peaks. This supports the idea that the transition is primarily an entropy driven transition. We can also get a good idea of the relative strengths of the transition for the different structures: The largest change occurs in the HTa$^-$ structure, and the smallest one happens in the LTa$^0$. 

In passing, we would like to emphasize that even though the occupancy histograms show that LaCoO$_3$ is in a mixed spin state, this is a spatially homogenous mixed state (such as those discussed in references [\onlinecite{Asai1998, Abbate}]) and not an inhomogeneous spin state, such as those discussed in various theoretical studies \cite{Zhang,Knizek,Zhuang,Kunes2011} and also discussed in relation to the Magnetic Circular Dichroism results in Ref. [\onlinecite{Haverfort}]. All of our calculations are performed using structures with only one crystallographically inequivalent Co ion. Thus, a spin state superstructure is beyond what our current calculations can reproduce. However, a very important conclusion of our DFT+eDMFT study is that a  \textit{ spatially inhomogenous mixed state is not necessary to reproduce the spin state and metal-insulator transitions in LaCoO$_3$}.

In this context, we would also like to point out that the possibility of a spatially inhomogenous spin state in which low- and intermediate- or high-spin Co$^{3+}$ ions form a spin state crystal is considered as a possible scenerio to explain the metamagnetic phase transitions observed under high magnetic fields.\cite{highfield1, highfield2,highfield3} A recent two-orbital DMFT study, on the other hand, favors an excitonic condensation scenerio.\cite{kunes_highfield} Our study does not address the effect of a magnetic field, nor does it cover the low-temperature range where the field induced transitions are observed. Nevertheless, our results underline the importance of both the intermediate spin and high spin states coexisting on the same ion, as well as the importance of charge fluctuations on the Co ion, which supports some of the findings of Ref. \onlinecite{Kunes_main} along with some of the explanations considered for the original high field experiments \cite{highfield4}.

\section{Contribution of Electronic Entropy} 

According to the entropy driven transition scenerio, which is supported by calorimetric measurements\cite{Stolen}, LaCoO$_3$ favors higher spin multiplets  at elevated temperatures because of the  associated gain in electronic entropy as a result of the high degeneracy of these high spin states - a point missed by first principles calculations at the level of DFT. Access to higher spin states is also made easier by a larger lattice constant due to the reduced crystal field splitting, so the gain in electronic entropy could also be a driving factor for the large thermal expansion seen in this material.

We calculated the contribution of the electronic entropy to the free energy using our state of the art DFT+DMFT implementation\cite{Birol}. 
In particular, we evaluated the Free Energy and the Electronic Entropy for both the 4K and 1143K structures (LTa$^-$ and HTa$^-$) at 1160K to predict if the structural changes make a considerable difference.\footnote{In our calculations, the phononic contribution to the entropy is not included since we employ the Born Oppenheimer approximation and the ion cores are considered to have well defined, stationary positions. While a first principles calculation of the phononic entropy is in principle possible, for example using the quasiharmonic approximation\cite{Dove_Book}, such an endeavour is very computationally demanding, and, to the best of our knowledge, has never been performed using an advanced first principles method such as DFT+DMFT. In this particular system, there is no indication of a particularly interesting soft phonon mode, and so ignoring the phononic entropy probably does not change any results in a significant way.}
The HTa$^-$ structure is indeed much higher in electronic entropy compared to the LTa$^-$ structure at 1160K and we observe the difference in $T\cdot S$ between these two structures to be $\sim$ 110 meV per formula unit. This unusually large difference emphasises the importance of electronic entropy to the transition.We also calculate the energy difference between the HTa$^-$ and LTa$^-$ structures to be $\sim$ 70 meV at 1160K with the LTa$^-$ being lower in energy. Thus we see that when the entropy is taken into account and the Free Energy (F=E-TS) is calculated, the high temperature structure HTa$^-$ becomes more stable purely due to the contribution of electronic entropy.  This result confirms the structural phase transition that is observed as a function of temperature. So, we can conclude that electronic entropy, which has been ignored in many first-principles studies of this material, is a leading factor in creating an anomalously large thermal expansion and driving the material to a high spin state.

\section{Summary} 
We studied the spin state transition of LaCoO$_3$ using state of the art fully charge self consistent DFT+ Embedded DMFT. By using different experimental and hypothetical crystal structures, we disentangled the effect of different components of the crystal structure and showed that both the thermal expansion and the presence of oxygen octahedral rotations have tremendous effect on the spin state transition of LaCoO$_3$. Our single site DMFT approach reproduced not only the spin state transition but also the intermediate phase which has nonzero magnetic moment but is insulating. This shows that the spin state and the metal-insulator transitions occur at different temperature scales and that the insulating phase with local magnetic moments can be reproduced without  necessarily involving cell doubling via mechanisms such as breathing distortions of spatially inhomogenous mixed spin states. Our results emphasize the importance of charge fluctuations on the Co ion due to hybridization with the O anions, and thus point to the inadequacy of a simple spin state picture with only one formal valence. We find that while the spin state transition is concurrent with a sudden change in occupation in the high spin multiplets and the metal-insulator transition with a jump in the intermediate spin probability, both low and intermediate spin states also have significant occupation in the whole temperature range. Finally, our work is the first calculation of the electronic entropy of LaCoO$_3$ and it points to the fact that the change in electronic entropy with temperature is significant and is large enough to drive the spin state transition in this material. 

\section{Acknowledgements} 
TB was supported by the Rutgers Center for Materials Theory. KH was supported by the NSF-DMR 1405303.

\appendix*
\section{Details of the DFT+DMFT Calculations}
We performed fully charge self-consistent  DFT+ Embedded DMFT calculations\cite{Dmft1}\cite{Dmft3}\cite{Dmft2} on LaCoO$_3$. Our implementation is based on the Wien2K all-electron DFT package.\cite{wien2k} For the DFT functional we use the GGA-PBE  functional. We employed a 10 atom unit cell for the 2 structures with rotations and 5 atom unit cells for the two structures without rotations and used 512 k points in the first Brillouin zone.    Our DMFT implementation uses state-of-the-art CTQMC\cite{haule} impurity solver based on the hybridization expansion (CTQMC-HYB) and all simulations are iterated to self-consistency. It is to be noted that our implementation makes use of projectors to embed/project out the impurity self-energy onto the lattice degrees of freedom. This prevents errors associated with downfolding using Wannier orbitals and allows us to achieve highly accurate charge self-consistency.\cite{Dmft3} We also account for octahedral rotations, wherever present, by applying local rotations so as to align our correlated orbitals with the local crystal field set up by the neighboring atoms.  In our simulations, we use a Hubbard U of 6.0 eV and Hund's coupling (J) of 0.7eV. We have also investigated the effects of varying the value of J, and found that this does not change the temperature at which the transitions occur, but merely changes the value of the observed $|S_z|$ by a small amount, with higher J values resulting in slightly larger values. We ignored spin-orbit interaction, as it is expected to be small in this 3d transition metal oxide. 

In our calculations, we included all d electrons of the Co ion in the impurity, performing the CTQMC calculation with all 5 orbitals. We used the Slater parametrization of the Coulomb U.\cite{dmft_coulombU}
The typical self-consistent calculation requires 30 self consistency cycles each with around 10 DFT iterations and one CTQMC iteration per cycle.  Our LDA+DMFT calculations gave us the Green's function ($G(i\omega)$) on the imaginary (Matsubara) axis which we analytically continued using the maximum entropy method to get the density of states on the real axis. 
The electronic entropy we calculate is the entropy of the impurity part, which includes all the d-shell electrons and is expected to be the dominant contribution.

Our implementation of DMFT is based on mapping the original lattice problem to an auxilliary impurity problem. We solve numerically exactly the impurity problem defined by the action:
\begin{widetext}  
\begin{multline}\label{equ:action}
\mathcal{S}=\int_0^\beta d\tau\psi^\dagger_{L\sigma}(\tau)\frac{\partial}{\partial\tau} \psi_{L\sigma}(\tau) +\int_0^\beta d\tau \int_0^\beta d\tau'\psi^{\dagger}_{L_1 \sigma}(\tau')\Delta_{L_1\sigma,L_2\sigma'} (\tau - \tau') \psi_{L_2 \sigma'}(\tau)\\
+\frac{1}{2}\int_0^\beta d\tau \sum_{L_1,L_2,L_3,L_4,\sigma,\sigma'} U_{L_1,L_2,L_3,L_4}\psi^\dagger_{L_1 \sigma}(\tau) \psi^\dagger_{L_2 \sigma'}(\tau) \psi_{L_3 \sigma'}(\tau) \psi_{L_4 \sigma}(\tau) 
\end{multline}

where $\tau$ is imaginary time, $L$ and $\sigma$ denote angular momentum and spin labels,$\psi$ and $\psi^\dagger$ are the annihilation and creation operators for impurity electrons and $\Delta$ is the hybridization function between our impurity and the bath, which encodes most of the lattice information. The on-site Coulomb repulsion between the Co d-electrons is given by
\begin{equation}
\hat{U}=\frac{1}{2}\sum_{L_1,L_2,L_3,L_4,\sigma,\sigma'} U_{L_1,L_2,L_3,L_4}c^\dagger_{L_1 \sigma} c^\dagger_{L_2 \sigma'} c_{L_3 \sigma'} c_{L_4 \sigma}
\end{equation}
 where
\begin{equation}
U_{L_1,L_2,L_3,L_4}=\sum_k \frac{4 \pi}{2k+1}\langle Y_{L_1}|Y_{km}|Y_{L_4}\rangle\langle Y_{L_2}|Y^*_{km}|Y_{L_3}\rangle F^k_{l_1,l_2,l_3,l_4}
\end{equation}
\end{widetext}
where $Y$ denotes spherical harmonics and $F^k$ denote Slater integrals. The CTQMC impurity solver gives us the impurity self-energy $\Sigma$. Since correlations are very local in real space, we embed this self-energy by expanding it in terms of quasi-localized atomic orbitals,$\ket{\phi^{m}_{n}}$;

\begin{equation}
\Sigma_{i\omega}(r,r')=\sum_{mm',nn'}\braket{r|\phi^{m}_{n}}\braket{\phi^{m}_{n}|\Sigma|\phi^{m'}_{n'}}\braket{\phi^{m'}_{n'}|r'}
\end{equation}

where $m,m'$ denote the sites of Co atoms in the most general cluster-DMFT implementation and $n,n'$ denote the atomic degrees of freedom for each site. Since, we perform single site DMFT, $\sum_{mm'}$ becomes $\delta_{m,m'}$. We then solve the Dyson equation in real space (or an equivalent complete basis such as the Kohn-Sham basis) according to the equation:

\begin{equation}
G_{i\omega}(r,r')=\left( \left(i \omega +\mu +\triangledown^{2} +V_{KS}(r) \right)\delta(r-r') - \Sigma_{i \omega}(r,r') \right)
\end{equation}

This procedure is what we define as embedded DMFT, because the self energy is embedded into a large Hilbert Space instead of constructing a Hubbard-like model by downfolding to a few bands using Wannier Orbitals. This method has the advantage that the correlations are much more localized in real space compared to the Wannier representation, which makes DMFT a much better approximation. In addition, this formulation of DFT+DMFT can be shown to be derivable from the  Luttinger Ward Functional which makes the formulation stationary and conserving \cite{Birol}.

\end{document}